\documentclass[11pt,a4paper]{article}
\usepackage[margin=1in]{geometry}
\usepackage{amsmath,amsfonts}
\usepackage{algorithmic}
\usepackage{algorithm}
\usepackage{array}
\usepackage{textcomp}
\usepackage{stfloats}
\usepackage{url}
\usepackage{verbatim}
\usepackage{graphicx}
\usepackage{subfigure}

\newcommand\blfootnote[1]{%
  \begingroup
  \renewcommand\thefootnote{}\footnote{#1}%
  \addtocounter{footnote}{-1}%
  \endgroup
}

\title{Analytic Model of Trans-axial Sensitivity in Cylindrical PET Systems Based on Solid Angle}

\author{
Boheng Lin\textsuperscript{c,\textdagger},\;
Zizhuo Xie\textsuperscript{b,\textdagger,*},\;
Bo Zhang\textsuperscript{c,*},\;
Lin Wan\textsuperscript{a},\;
Ao Qiu\textsuperscript{c},\;
and Qingguo Xie\textsuperscript{b,c}
\\[0.8em]
\normalsize
\begin{minipage}{0.85\textwidth}\centering
\textsuperscript{a}School of Software Engineering, Huazhong University of Science and Technology, Wuhan, China\\[2pt]
\textsuperscript{b}School of Information Science and Technology, University of Science and Technology of China, Hefei, China\\[2pt]
\textsuperscript{c}Department of Biomedical Engineering, Huazhong University of Science and Technology, Wuhan, China
\end{minipage}
}
\date{}

\begin{document}
\maketitle
\blfootnote{\textsuperscript{\textdagger}These authors contributed equally to this work.}
\blfootnote{\textsuperscript{*}Corresponding authors: z8xie@ucsd.edu, bozhang-pet@hust.edu.cn}

\begin{abstract}
In positron emission tomography (PET), a clear theoretical model describing how system sensitivity varies as a source is moved trans-axially within the field of view (FOV) is lacking. The current understanding and practical intuition often suggest that sensitivity is maximum at the center of the FOV, an assumption reflected in standardized protocols. In this work, we derive an analytic model for the trans-axial-plane sensitivity distribution in a cylindrical PET scanner based on solid angle. The model, formulated as a function of trans-axial offset from the center, is validated through both Monte Carlo simulations and physical experiments on a representative system. We find that the derived theoretical distribution is essentially consistent with simulation and experimental results, revealing a non-intuitive feature: sensitivity increases with trans-axial offset, peaks at the edge of the FOV, and drops off sharply beyond it. This study provides the first closed-form model of trans-axial geometric sensitivity in cylindrical PET scanners, offering a vital benchmark for isolating detector technology improvements and revealing a non-intuitive, offset-dependent sensitivity profile that enables new protocol optimization strategies.
\end{abstract}




\section{Introduction}
Positron Emission Tomography (PET) is widely used for diagnosing and monitoring cancer, neurological disorders, and cardiovascular conditions\cite{Li2017PET, Farwell2014PET,Politis2012PET, 9353691}. Among the many metrics that characterize PET performance, sensitivity is one of the most critical~\cite{nguyen2015digitalPET, Calderon2023, BUDINGER1998247}. Standardized protocols such as NEMA NU 2-2018\cite{nema2018} define sensitivity measurement procedures at the center of the field of view (CFOV) and at a trans-axial offset of 10 cm. This convention, used in many system performance studies\cite{jakoby2011performance, chen2020performance, bonifacio2023analytical, chicheportiche2020comparison}, could suggest that sensitivity is either maximized or sufficiently characterized near the CFOV.

However, empirical studies present inconsistent findings. Some report higher sensitivity at 10 cm than at the CFOV\cite{chen2020performance, chicheportiche2020comparison}, others report the opposite\cite{jakoby2011performance}, and a few describe non-monotonic trends across the trans-axial plane\cite{visser2009inveon}.

Empirical inconsistencies in reported trans-axial sensitivity underscore the absence of a theoretical baseline. While Bonifacio et al. provided an analytical characterization of sensitivity along the axial dimension \cite{bonifacio2023analytical}, a corresponding quantitative description for the trans-axial plane—that is, an analytical model of sensitivity as a function of trans-axial source position—remains absent in the literature.

In this study, we present a closed-form analytical model derived from the solid angle subtended by a source to characterize the trans-axial-plane sensitivity distribution in a cylindrical PET scanner. This theoretical formulation is subsequently validated using Monte Carlo simulations and physical experiments, establishing its applicability beyond idealized conditions.

\section{Methods}
\subsection{Analytical Model}
\subsubsection{Definition of Sensitivity}
The sensitivity of a cylindrical PET scanner is defined as the ratio of recorded true coincidence event rate to the activity of the positron-emitting source. Sensitivity is primarily determined by three factors: the detection efficiency at 511 keV, the detector packing fraction, and the solid angle subtended by the detector geometry\cite{CherryDahlbomPETPhysics, amoyal2021development, BUDINGER1998247, Madsen2004, nema2018, ERIKSSON2007836}. Among these, the present analytical model focuses exclusively on the contribution of the solid angle, often called geometric sensitivity, which is governed by the angular coverage of the detector walls as viewed from the radiation source location\cite{PerezBenito2023, Ficke1996SpheroidPET}. Specifically, the ratio of recorded true coincidence events to the source's radioactivity can be expressed in terms of the fraction of the total solid angle of the space subtended by the detector walls; let the geometric sensitivity be denoted as \(S_{\mathrm{g}}\):

\begin{equation}
S_{\mathrm{g}} := \frac{\text{solid angle subtended by detector walls}}{4\pi}
\end{equation}

For a positron-electron annihilation event to be classified as a true coincidence, both of the resulting 511 keV gamma photons must be detected by opposing detectors within a defined coincidence time window and energy window\cite{Vallabhajosula2023,CherryDahlbomPETPhysics}. In a cylindrical PET system, the axial ends of the scanner are open circular caps, hereafter referred to as Cap 1 and Cap 2. A positron-electron annihilation event is classified as a true coincidence if and only if both resulting 511 keV gamma photons are retained within the detection volume—that is, neither photon exits the system through Cap 1 or Cap 2.

The probability of a gamma photon escaping through one of the caps before interacting with the detector walls is determined by the solid angle subtended by the caps at the position of the radiation source\cite{Camborde2001}. Denote this solid angle as \(\Omega\). Then the geometric sensitivity for coincidence events in a cylindrical PET system can be expressed as:
\begin{equation}
S_{\mathrm{g}} = (4\pi - \Omega) / 4\pi = 1 - \Omega / 4\pi.
\end{equation}

\subsubsection{Geometric Modeling}
We set up the geometric analysis within the following spatial framework:
\begin{itemize}
    \item Define the origin as \((0, 0, 0)\).
    \item Define the \(Z\)-axis as the horizontal axis.
    \item Define the \(X\)-axis as the vertical axis.
    \item Define the \(Y\)-axis perpendicular to both the \(Z\)-axis and the \(X\)-axis, completing a right-handed coordinate system.
\end{itemize}

The cylindrical PET system is oriented so that:
\begin{itemize}
    \item Its axial direction aligns with the \(Z\)-axis.
    \item Its cross-sectional planes are parallel to the \(XY\)-plane, representing slices of the cylinder.
    \item The cylinder extends from \(z = -L'\) to \(z = L'\) along the \(Z\)-axis, and it has a constant radius \(R'\) in the \(XY\)-plane. It can be represented by the equation:
    \begin{equation}
    x^2 + y^2 = R'^2, \quad -L' \leq z \leq L'.
    \end{equation}
\end{itemize}

The spatial position of the radiation source is represented by a point \(P(r', y', L' - h')\), where:
\begin{itemize}
    \item \(r'\) and \(y'\) specify the source’s position within the \(XY\)-plane (cylindrical cross-section).
    \item \(L' - h'\) specifies the source’s position along the axial \(Z\)-axis, where \(h'\) is the \(Z\)-axis distance between \(z = L'\) and the z-coordinate of point \(P\).
\end{itemize}

Since \(L'\), \(R'\), \(r'\), and \(h'\) are chosen arbitrarily, we can normalize to simplify later operations. Let:
$R = R'/R' = 1\qquad L = L'/R'\qquad r = r'/R'\qquad h = h'/R'.$
The cylinder can now be represented as:
\begin{equation}
x^2 + y^2 = 1^2, \quad -L \leq z \leq L.
\end{equation}
And the point \(P\) is now \(P(r, y, L - h)\).

\begin{figure*}[htbp]
    \centering
    \subfigure[]{
        \includegraphics[width=0.40\linewidth]{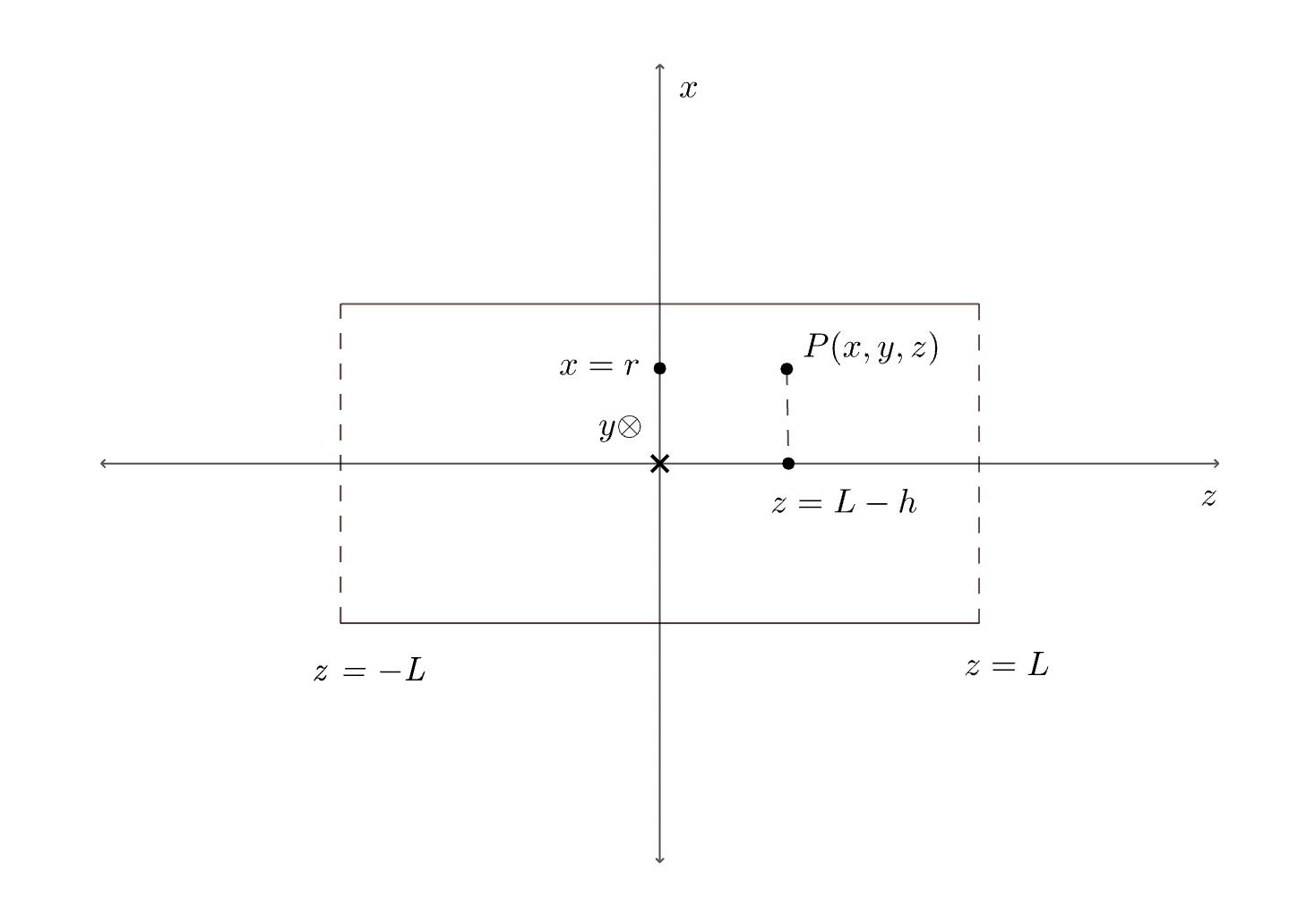}
    }
    \hspace{0.01\linewidth}
    \subfigure[]{
        \includegraphics[width=0.35\linewidth]{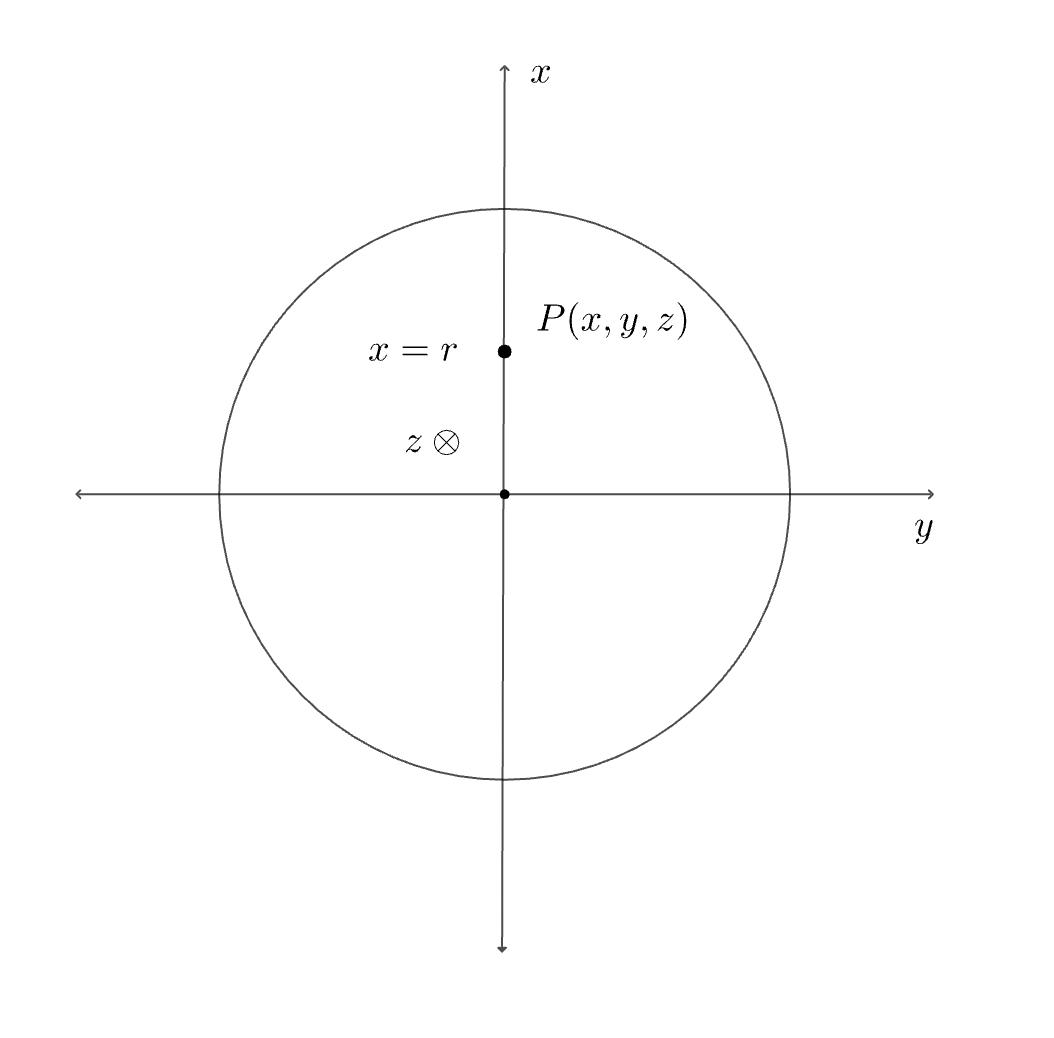}
    }
    \caption{Diagrams showing the PET cylindrical system within the defined spatial framework: (a) view sliced along the axial direction, and (b) view sliced along the trans-axial direction.}
    \label{fig:PET_setup}
\end{figure*}

This has been visually summarized in Figure~\ref{fig:PET_setup}.

\subsubsection{Computing Solid Angle}
The geometric sensitivity \( S_{\mathrm{g}} \) is derived from the solid angle \( \Omega \) subtended by the two detector endcaps of the cylindrical PET scanner. For a point source located at a trans-axial offset \( r \) and axial position \( z \), \( \Omega \) is obtained by integrating the differential solid angle over the cap surfaces.  

Using the geometric relation \( d\Omega = h\, dA / \rho^3 \), where \( h \) is the perpendicular distance from the source to the cap plane and \( \rho \) is the distance from the source to a surface element \( dA \), the solid angle can be expressed as:
\begin{equation}
    \Omega = \iint\limits_{H} \frac{h \, dy \, dx}{\left[(x-r)^2 + y^2 + h^2\right]^{3/2}},
\end{equation}

where the integration domain \( H \) corresponds to the area of one endcap.

Applying Green’s theorem~\cite{01630569708816786}, the surface integral is transformed into a line integral along the boundary \( \partial H \) of the cap:
\begin{equation}
    \Omega = \oint\limits_{\partial H} \frac{-h y \, dx}{\left[(x-r)^2 + h^2\right] \sqrt{(x-r)^2 + y^2 + h^2}}.
    \label{eq:omega}
\end{equation}

Finally, the geometric sensitivity is given by:
\begin{equation}
    S_{\mathrm{g}} = 1 - \frac{\Omega}{2\pi},
\end{equation}
where $\Omega$ is given in Equation~\eqref{eq:omega}.

A detailed derivation of the above steps is provided in appendix~\ref{app:solid_angle_dev}.
\subsubsection{Geometric Case Analysis}
In order to derive the analytic expression for geometric sensitivity, we consider
the spatial relationship between two spherical caps, denoted as Cap 1 and Cap \(2'\) (see appendix~\ref{app:case_analysis}).
The union of these regions exhibits distinct geometric configurations depending on
the relative position and size of Cap \(2'\) with respect to Cap 1. Accordingly,
the analysis is divided into the following two cases:

\begin{enumerate}
    \item When Cap \(2'\) is entirely contained within Cap 1, that is, \(m + R_2 \leq R\),
    the union of the two regions is reduced simply to Cap 1.
    \item When Cap \(2'\) extends beyond the boundary of Cap 1, i.e., \(m + R_2 > R\),
    the union comprises the entirety of Cap 1 together with the portion of Cap \(2'\)
    that protrudes outside Cap 1.
\end{enumerate}

Under the first configuration, the geometric sensitivity can be expressed as
\begin{equation}
    \begin{aligned}
        S_{\mathrm{g}} &= 1 - \frac{1}{2\pi} \int_{0}^{2\pi} \frac{h}{\big[(\cos \theta - r)^2 + \sin^2 \theta + h^2\big]^{1/2}}
        \cdot \frac{\sin^2 \theta\, \mathrm{d} \theta}{(\cos \theta - r)^2 + h^2}.
    \end{aligned}
    \label{eq:Sg1}
\end{equation}

For notational simplicity, we denote the integrand in Equation~\eqref{eq:Sg1} as
\[
F(\theta, r, h, L),
\]
where \(F(\cdot)\) captures the complete dependence of the expression on the relevant parameters.

In the second configuration, wherein Cap \(2'\) extends beyond the boundary of Cap 1, the
geometric sensitivity involves contributions from both the overlapping and protruding
regions. In this case, the geometric sensitivity is given by
\begin{equation}
    \begin{aligned}
        S_{\mathrm{g}} =&\ 1 - \frac{1}{2\pi} \int_{\phi_1}^{2\pi - \phi_1} F(\theta, r, h, L)\, \mathrm{d}\theta \\
        &\quad - \frac{1}{2\pi} \int_{-\pi + \phi_2}^{\pi - \phi_2}
        \frac{h}{\big[(R_2 \cos \theta + m - r)^2 + R_2^2 \sin^2 \theta + h^2\big]^{1/2}}
        \cdot \frac{R_2^2 \sin^2 \theta\, \mathrm{d}\theta}{(R_2 \cos \theta + m - r)^2 + h^2}.
    \end{aligned}
    \label{eq:Sg2}
\end{equation}

To streamline notation, the integrand appearing in the second integral of Equation~\eqref{eq:Sg2} is
denoted as
\[
G(\theta, r, h, L, R_2),
\]
which encapsulates the parameter dependence of the corresponding geometric contribution.

A complete and detailed derivation of these expressions is provided in appendix~\ref{app:case_analysis}.

Consolidating the two cases into a single result, where the general geometric sensitivity of the system can be modeled by the following distribution:
\begin{equation}
  S_g(r,h) =
\begin{cases}
1 - \dfrac{1}{2\pi} \displaystyle\int_{0}^{2\pi} F(\theta, r, h, L) \, d\theta, & m + R_2 \leq R \\[8pt]
\begin{aligned}
1 - \dfrac{1}{2\pi} \Bigg( & \displaystyle\int_{\phi_1}^{2\pi-\phi_1} F(\theta, r, h, L) \, d\theta \\
& + \displaystyle\int_{-\pi+\phi_2}^{\pi-\phi_2} G(\theta, r, h, L, R_2) \, d\theta \Bigg),
\end{aligned} & m + R_2 > R
\end{cases}
\label{eq:Sg}
\end{equation}

\(S_g(r,h)\) gives the geometric sensitivity distribution as a results of a point source, to obtain the corresponding distribution for a line source, we integrate along the axial direction as follows:
\begin{equation}
    S_{g, \,line}(r) = \int_0^LS_g(r,h) \, dh
    \label{eq:Sg_line}
\end{equation}
This is done in order to match the simulation and experimental conditions presented in the following section.

\subsection{Experiments and Simulations}

\subsubsection{GATE Simulations}
Simulations were performed in GATE\cite{SJan_2004} using a PET scanner with a radius \( R = 403.8 \) mm and an axial length \( 2L =306 \) mm. The scanner used lutetium yttrium osmium (LYSO) scintillator crystals, arranged in a \( 6 \times 6 \) grid within each detector block. Each crystal had dimensions of \( 20.0 \, \text{mm} \times 3.9 \, \text{mm} \times 3.9 \, \text{mm} \).

The system consists of three detector rings, each ring containing 48 detector modules, totaling 144 modules. Each detector module contains \( 1 \times 2 \times 4 \) sub-blocks, with dimensions \( 20.0 \, \text{mm} \times 25.5 \, \text{mm} \times 25.5 \, \text{mm} \).

A cylindrical \( ^{18}\text{F} \) radioactive source with a radius of \( 0.01 \) mm and a length of \( 700 \) mm was used. The source had an activity of \( 7.4 \) MBq (\( 7.4 \times 10^6 \) Bq). Sensitivity values were obtained at trans-axial offsets in increments of \( 25 \) mm, covering a range from \( 0 \) mm to \( 400 \) mm, resulting in a total of 17 simulation data points. These values were calculated in accordance with the NEMA standard~\cite{nema2018}.

The following thresholding methods were applied:

\begin{itemize}
    \item \textbf{Energy Window Filtering:} Only events with energy levels within the window of \([420, 650] \, \text{keV}\) were considered valid.
    \item \textbf{Time Window Filtering:} Only photon pairs that were detected within 4 ns of each other were considered true coincidence events.
    \item \textbf{Distance Thresholding/FOV:} Events detected outside the FOV radius were discarded. Simulations were conducted for 500 mm FOV diameter, 600 mm FOV diameter, as well as no FOV restriction.
\end{itemize}

\subsubsection{Experiment}
Experiments were conducted using the DigitMI 930 scanner (RAYSOLUTION Healthcare Co., Ltd.)~\cite{Zhang2025DigitMI930}, whose dimensions matched those employed in the simulations.

A cylindrical \textsuperscript{18}F radioactive line source, with a radius of 0.01~mm and a length of 700~mm, was prepared at an activity of 7.4~MBq. This source was positioned at successive trans-axial offsets ranging from 0~mm to 375~mm in 25~mm increments, yielding 16 distinct sensitivity measurements. Data acquisition and subsequent sensitivity calculations adhered to the NEMA standard~\cite{nema2018}, ensuring methodological consistency with established benchmarks.

The following thresholding methods were applied:  
\begin{itemize}
    \item \textbf{Energy Window Filtering:} Only events with energy levels within the window of \([420, 650] \, \text{keV}\) were considered valid. 
    \item \textbf{Time Window Filtering:} Only photon pairs detected within 4 ns of each other were considered true coincidence events.  
    \item \textbf{Distance Thresholding/FOV:} A 500 mm FOV diameter was employed, discarding events detected beyond this range.  
\end{itemize}

\section{Results}
\begin{figure}[h]
     \centering
     \includegraphics[width=3.6in]{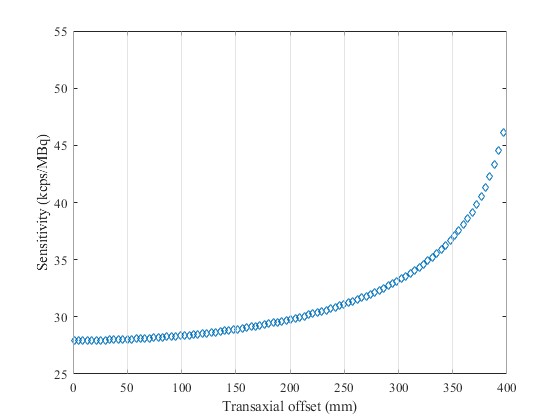}
     \caption{Plot of the geometric sensitivity as a function of trans-axial offset, calculated based on Equation~\eqref{eq:Sg_line}. The plot considers 100 evenly spaced points within \([0, R]\) for a PET system with dimensions \(R = 405\) mm and \(L = 153\) mm.}
     \label{fig:Geo_res}
\end{figure}

\begin{figure*}[h]
    \centering
    \subfigure[Simulation with FOV = 500 mm]{
        \includegraphics[width=0.30\linewidth]{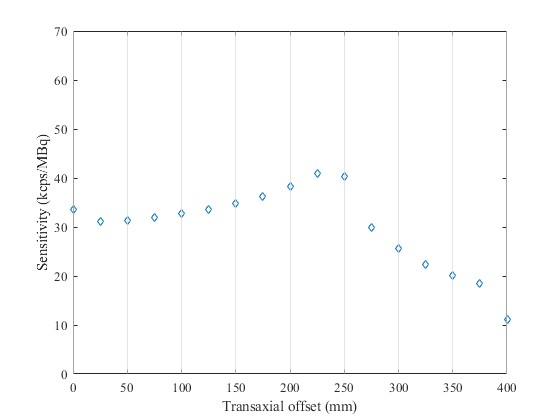}
    }
    \hspace{0.01\linewidth}
    \subfigure[Simulation with FOV = 600 mm]{
        \includegraphics[width=0.30\linewidth]{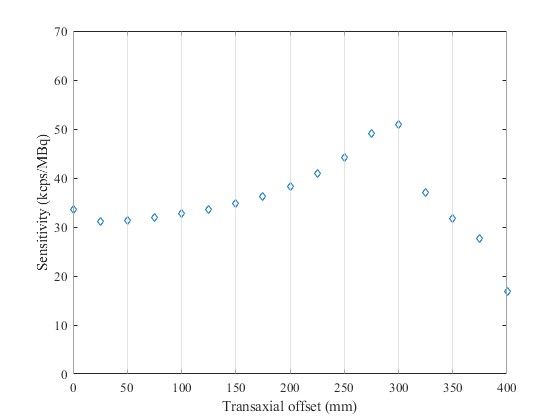}
    }
    \hspace{0.01\linewidth}
    \subfigure[Simulation with no FOV]{
        \includegraphics[width=0.30\linewidth]{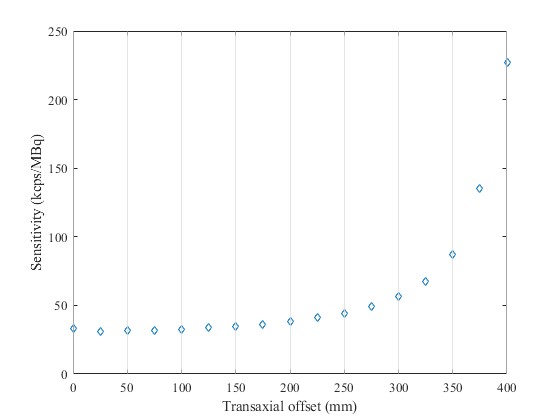}
    }
    \caption{GATE simulated distributions of sensitivity with respect to trans-axial offset for FOV diameter of (a) 500 mm, (b) 600 mm, and (c) no FOV.}
    \label{fig:Sim}
\end{figure*}

\begin{figure}[h]
     \centering
     \includegraphics[width=0.4\textwidth]{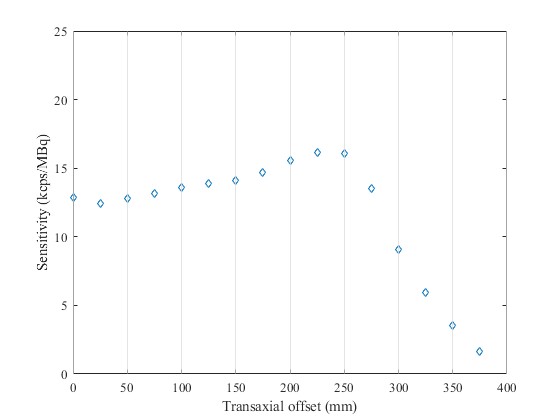}
     \caption{Experimental sensitivity w.r.t trans-axial offset. The effective FOV of the system was set to 500mm for this experiment.}
     \label{fig:Ex_rad}
\end{figure}

For the analytical model, Eq.~\eqref{eq:Sg_line} was evaluated at 100 evenly spaced trans-axial positions over the range 0–400~mm, and the resulting sensitivity distribution is shown in Figure~\ref{fig:Geo_res}. The corresponding sensitivity distributions obtained from simulations with different FOV configurations (500~mm, 600~mm, and no FOV) are presented in Figure~\ref{fig:Sim}, while the experimentally measured sensitivity values are shown in Figure~\ref{fig:Ex_rad}. Quantitative sensitivity values from the analytical model and simulations at trans-axial offsets from 0~mm to 400~mm in increments of 25~mm, together with experimental measurements from 0~mm to 375~mm, are summarized in Table~\ref{tab:total}.

\begin{figure*}[h]
    \centering

    \subfigure[Simulation (FOV = 500 mm)]{
        \includegraphics[width=0.30\linewidth]{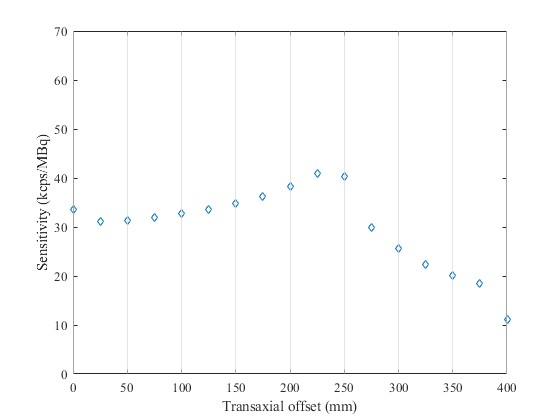}
    }
    \hspace{0.01\linewidth}
    \subfigure[Experimental (FOV = 500 mm)]{
        \includegraphics[width=0.30\linewidth]{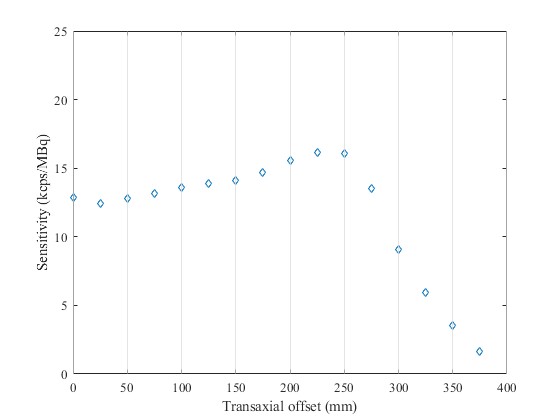}
    }
    \hspace{0.01\linewidth}
    \subfigure[Analytical]{
        \includegraphics[width=0.30\linewidth]{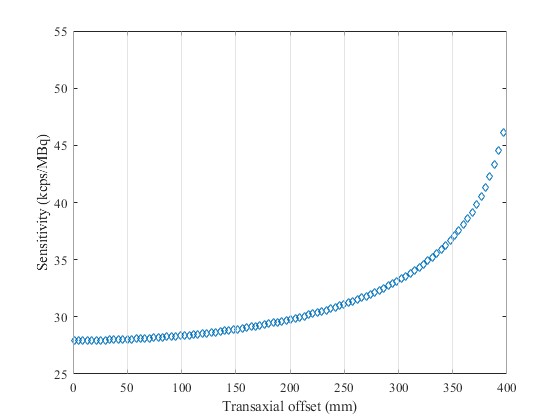}
    }

    \vspace{0.02\linewidth}

    \subfigure[Simulation (FOV = 600 mm)]{
        \includegraphics[width=0.30\linewidth]{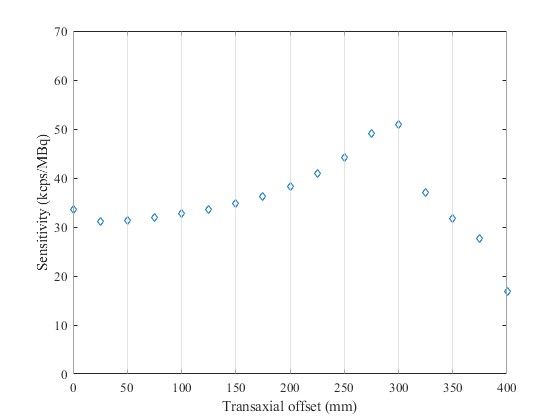}
    }
    \hspace{0.01\linewidth}
    \subfigure[Experimental (FOV = 500 mm)]{
        \includegraphics[width=0.30\linewidth]{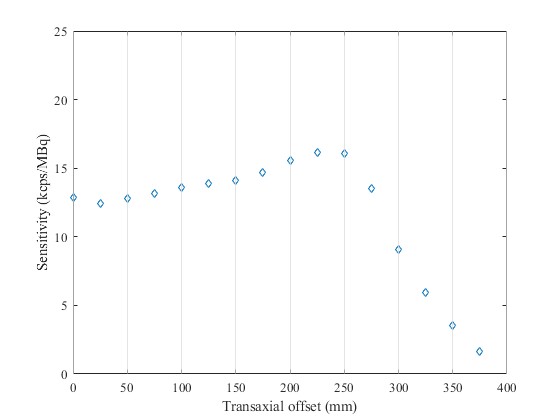}
    }
    \hspace{0.01\linewidth}
    \subfigure[Analytical]{
        \includegraphics[width=0.30\linewidth]{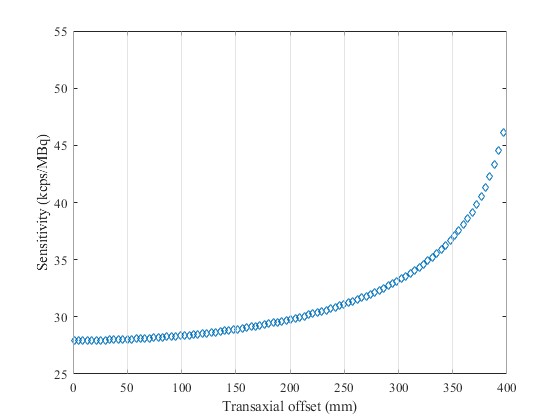}
    }

    \vspace{0.02\linewidth}

    \subfigure[Simulation (no FOV)]{
        \includegraphics[width=0.30\linewidth]{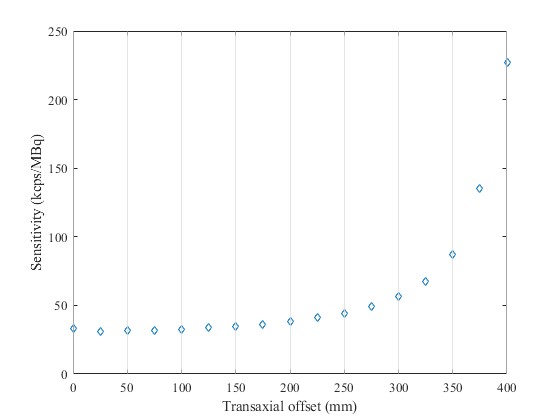}
        \label{subfig:noFOV}
    }
    \hspace{0.01\linewidth}
    \subfigure[Experimental (FOV = 500 mm)]{
        \includegraphics[width=0.30\linewidth]{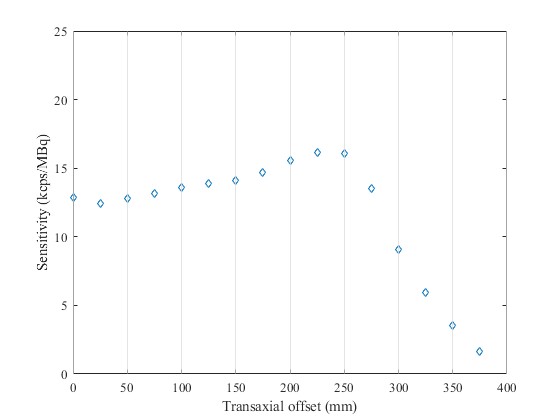}
    }
    \hspace{0.01\linewidth}
    \subfigure[Analytical]{
        \includegraphics[width=0.30\linewidth]{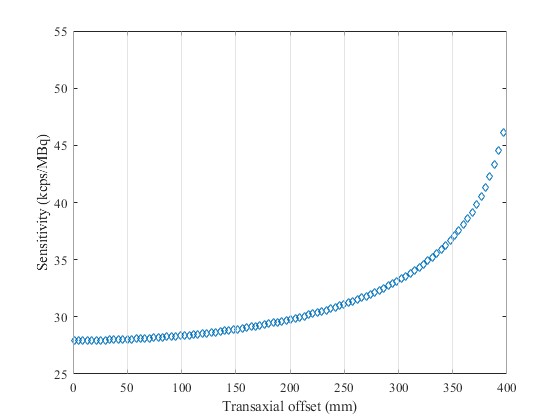}
    }

    \caption{Comparison of sensitivity as a function of trans-axial offset obtained from GATE simulations (left column), experimental measurements (middle column), and analytical predictions (right column).}
    \label{fig:Sim_Exp_Ana}
\end{figure*}

As shown in Figure~\ref{fig:Sim_Exp_Ana}, all three data sources exhibit an initial increase in sensitivity with increasing trans-axial offset. For the analytical model, the sensitivity increases monotonically across the entire trans-axial range. A qualitatively identical trend is observed in the simulation performed without a FOV constraint (Figure~\ref{subfig:noFOV}), where sensitivity continues to increase rapidly at large offsets due to the absence of event rejection. In contrast, both the experimental measurements and simulations employing a finite FOV demonstrate a distinct rise-and-fall behavior: following an initial increase, the sensitivity reaches a maximum and subsequently decreases as the trans-axial offset increases further. The location of this peak is governed by the configured FOV size, occurring at approximately half of the FOV diameter (around 250~mm for the 500~mm FOV and 300~mm for the 600~mm FOV). Beyond these offsets, events originating outside the FOV are progressively rejected, leading to a reduction in measured sensitivity.

Despite differences in absolute magnitude, the internal trends across experiments, simulations, and the analytical model are consistent and physically intuitive. The simulations correctly reproduce the experimentally observed peak-and-drop-off behavior when a finite FOV is applied, while the analytical model captures the underlying geometrical sensitivity trend in the absence of FOV constraints. These results confirm that, under idealized geometric conditions, sensitivity is not necessarily maximized at the CFOV but instead increases with trans-axial offset, reaching a maximum at the boundary imposed by the FOV or, in the absence of such constraints, at the physical limits of the scanner geometry.

\begin{table}[h]
\centering
\caption{Calculated sensitivity values for experiment, simulations, and analytical model. Sensitivity values are reported in kcps/MBq.}
\label{tab:sensitivity_vs_offset}
\small
\setlength{\tabcolsep}{3pt}
\renewcommand{\arraystretch}{1.15}

\makebox[\columnwidth][c]{%
\begin{tabular}{c c c c c c}
\hline
\textbf{Offset (mm)} &
\textbf{Experiment} &
\textbf{Sim. (500 mm)} &
\textbf{Sim. (600 mm)} &
\textbf{Sim. (No FOV)} &
\textbf{Model} \\
\hline
0   & 12.86 & 33.49 & 33.49 & 33.49 & 27.94 \\
25  & 12.44 & 31.04 & 31.04 & 31.04 & 27.96 \\
50  & 12.81 & 31.38 & 31.38 & 31.38 & 28.04 \\
75  & 13.15 & 31.92 & 31.92 & 31.92 & 28.16 \\
100 & 13.56 & 32.68 & 32.68 & 32.68 & 28.35 \\
125 & 13.90 & 33.63 & 33.63 & 33.63 & 28.59 \\
150 & 14.12 & 34.71 & 34.71 & 34.71 & 28.90 \\
175 & 14.71 & 36.32 & 36.32 & 36.32 & 29.29 \\
200 & 15.59 & 38.33 & 38.33 & 38.33 & 29.77 \\
225 & 16.13 & 41.01 & 41.01 & 41.01 & 30.37 \\
250 & 16.09 & 40.25 & 44.27 & 44.27 & 31.10 \\
275 & 13.55 & 29.98 & 49.12 & 49.12 & 32.02 \\
300 & 9.09  & 25.60 & 50.82 & 56.21 & 33.19 \\
325 & 5.97  & 22.44 & 37.12 & 67.24 & 34.74 \\
350 & 3.54  & 20.18 & 31.71 & 87.14 & 36.91 \\
375 & 1.65  & 18.57 & 27.70 & 135.26 & 40.34 \\
400 & --    & 11.18 & 16.94 & 226.75 & 47.88 \\
\hline
\end{tabular}}
\label{tab:total}
\end{table}

\section{Discussion and Future Work} 
In this study, we derived an analytical model for the trans-axial sensitivity distribution of a cylindrical PET scanner based purely on solid angle considerations. The model revealed the non-intuitive feature that sensitivity increases with trans-axial offset. This trend was corroborated by simulations and experiments, with the simulation performed without a FOV constraint matching the shape of the analytical model. For configurations employing a finite FOV, both simulations and experiments confirmed that sensitivity follows the same increasing trend \textit{within} the FOV but drops sharply beyond it, as expected, because the FOV defines the trans-axial acceptance range for valid coincidence events.

The formulation of this analytical model carries significant implications for both theoretical understanding and practical application. First, it effectively decouples geometric contributions from detector-dependent factors in sensitivity analysis. This distinction is useful, as it establishes a geometry-only reference against which future technological innovations, such as novel detector materials, photodetectors, or readout electronics, can be evaluated. By isolating geometric effects, the model provides a normalization framework that facilitates clearer attribution of performance improvements to genuine advances in detector technology rather than to variations in scanner geometry.

Second, the model introduces an additional degree of freedom for system optimization under constrained imaging conditions. In low-dose or time-limited imaging protocols, where sensitivity is often the limiting factor, the ability to increase sensitivity through deliberate trans-axial offset of the source or patient presents a practical trade-off. Clinicians and system operators could, in principle, choose to offset the subject within the FOV to enhance count statistics, albeit at the cost of workflow convenience and potential repositioning effort. This possibility underscores the model’s utility in guiding scenario-specific protocol adjustments.

It should be noted that while the analytical model challenges the conventional assumption that sensitivity is maximized at the center of the CFOV, clinical and experimental protocols often retain CFOV positioning for reasons beyond sensitivity alone. Central placement is frequently preferred for workflow simplicity, patient comfort, and alignment with system isocenter, which simplifies reconstruction and calibration. Moreover, at larger trans-axial offsets, parallax error and spatial resolution degradation become increasingly pronounced, which may outweigh the benefits of enhanced sensitivity in terms of overall image quality~\cite{Lewellen2004PETSystems}. Thus, the model does not invalidate current practices but rather enriches the understanding of the underlying sensitivity landscape, enabling more informed decisions when trade-offs between sensitivity, workflow, and image quality are considered.

With regard to the comparisons made in this study , it should be noted that they are primarily qualitative. The overall system sensitivity is a result of three principal factors: detector efficiency at 511keV, detector packing fraction, and geometric solid angle. Our analytical model accounts only for the solid angle, whereas the simulations and experiments inherently incorporate all three. Consequently, while the qualitative agreement validates the geometric component, a fully rigorous quantitative comparison requires integration of the remaining factors.

Future work will focus on incorporating the two omitted components: detector efficiency and packing fraction. The packing fraction, defined as the ratio of active detector solid angle coverage to the total solid angle subtended by the PET cylinder, accounts for gaps between detector crystals and modules. As this factor is expected to be approximately constant across the trans-axial plane, it can be modeled by a simple multiplicative scaling of the current analytical result. 

Modeling detector efficiency is more complex. The coincidence intrinsic efficiency is given by
\begin{equation}
    \varepsilon^2 = (1-e^{-\mu d})^2 \times \Phi^2,
\end{equation}
where \( \mu \) is the linear attenuation coefficient of the detector material, \( d \) is the gamma photon path length within the crystal before escaping, and \( \Phi \) is the photopeak fraction within the selected energy window~\cite{CherryDahlbomPETPhysics}. The critical variable here is the path length \( d \). For a source at the center of the scanner, photons typically enter crystals normally, making \( d \) roughly equal to the crystal length. However, as the trans-axial offset increases, the average incidence angle changes, and photons may traverse multiple crystals, altering the effective \( d \). The primary challenge, therefore, is to derive the functional dependence of this path length on trans-axial offset, \( d(r) \), to complete a comprehensive analytical sensitivity model.

\section{Conclusion}
This study presents the first closed-form analytical model for trans-axial sensitivity in cylindrical PET scanners, based on geometric solid angle. The model demonstrates that sensitivity increases with trans-axial offset, peaking at the scanner’s manually configured or physical edge, a finding that contradicts the common assumption of maximum sensitivity at the center. Simulations without a field-of-view (FOV) restriction match the theoretical trend, while experiments and FOV-limited simulations confirm that sensitivity peaks at the FOV boundary before declining.

The work establishes an analytical model for the geometric sensitivity in cylindrical PET, providing a theoretical baseline for sensitivity analysis. Future efforts will integrate detector efficiency and packing fraction to develop a complete analytical sensitivity framework.

\section*{Acknowledgments}
The authors gratefully acknowledge the support provided by RAYSOLUTION Healthcare Co., Ltd. for this work. The authors also extend their sincere appreciation to the PET Center of Union Hospital, Tongji Medical College, Huazhong University of Science and Technology, for the generous provision of radioisotopes and related resources.

\appendix
\section{Appendix: Computing Solid Angle}
\label{app:solid_angle_dev}
The computation of \(S_{\mathrm{g}}\) reduces to determining \(\Omega\), which is obtained by integrating the differential solid angle over the area of the caps. The differential solid angle is given by:
\begin{equation}
d\Omega = \cos(\theta) \, dA / \rho^2,
\end{equation}
where \(dA\) is the infinitesimal surface area element of on the caps, \(\theta\) is the angle between the vector connecting the source point \(P\) to \(dA\) and the surface normal, and \(\rho\) is the distance between \(P\) and \(dA\). Substituting \(\cos(\theta) = h / \rho\), where \(h\) is the perpendicular distance from \(P\) to \(dA\), we can rewrite \(d\Omega\) as:
\begin{equation}
d\Omega = h \, dA / \rho^3.
\end{equation}

\begin{figure}[!t]
     \centering
     \includegraphics[width=0.45\textwidth]{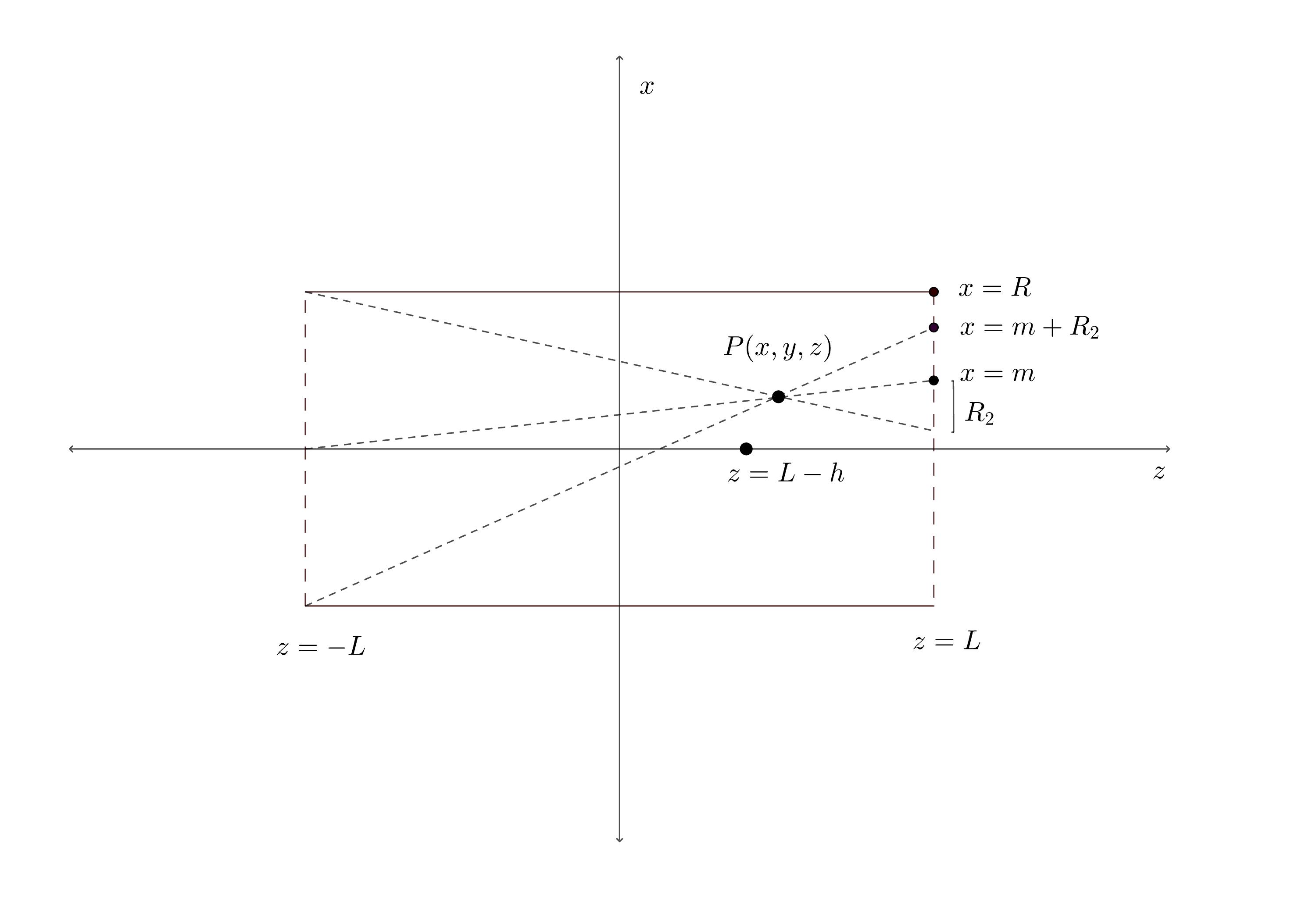}
     \caption{Diagram showing the projection of Cap 2 (left cap) onto Cap 1 (right cap) through the point \(P\), to form Cap \(2'\).}
     \label{fig:Projection}
\end{figure}

Since the system exhibits reflectional symmetry about the XY-plane, it is sufficient to analyze only the right half of the geometry. Without loss of generality, we therefore restrict attention to the region bounded by:
$x^2 + y^2 \leq R^2, \quad 0 \leq z \leq L.$
To account for the omitted left half, the left end cap (Cap 2) is projected onto the plane of the right end cap (Cap 1) located at \( z = L \), resulting in the projected cap, denoted as Cap \(2'\), as illustrated in Figure~\ref{fig:Projection}.

The projected Cap \( 2' \) will have radius \(R_2\):
$R_2 = R (L - h) / (L + h),$
The center of Cap \( 2' \) will be a distance \( m \) from the center of Cap 1, where:
$m = r \cdot 2L / (2L - h).$ The total solid angle \( \Omega \) is the union of the solid angles subtended by Cap 1 (\( \Omega_1 \)) and Cap \( 2' \) (\( \Omega_2' \)):
$\Omega = \Omega_1 \cup \Omega_2'.$

Furthermore, due to the circular symmetry of the system within the XY-plane, it is sufficient to consider trans-axial offsets along the x-axis only. This allows us to disregard the y-coordinate and simplify the source position to the point
\(P(r, 0, L - h).
\)

Now, consider the integration of \(d\Omega\). The distance \( \rho \) from a point \((x, y, L)\) on the plane \(z = L\) in the system to \( P \) is given by:
$\rho = [(x - r)^2 + y^2 + h^2]^{1/2}.$ To find \(\Omega\), we integrate \(d\Omega\) over the total area of interest, call it \(H\):
\begin{equation}
    \Omega = \iint \limits_{H} \frac{h dA}{\rho^3} = \iint \limits_{H} \frac{h dy dx}{[(x - r)^2 + y^2 + h^2]^{3/2}}
\end{equation}

We can turn this integral over an area into a line integral over the boundaries of the area by using Green's Theorem\cite{01630569708816786} and the fact that: 
\begin{equation}
    \int \frac{\mathrm{d} x}{(x^2 + k^2)^\frac{3}{2}} = \frac{x}{k^2 \cdot \sqrt{x^2 + k^2}} + C
\end{equation}

Substituting the integral variable of the above formula into $y$, taking $C = 0$, and applying Green's Formula, we have:
\begin{equation}
    \begin{aligned}
         \Omega &= \iint \limits_{H} \frac{-h \mathrm{d} y \mathrm{d} x}{[(x - r)^2 + y^2 + h^2]^{3/2}} \\
         &= \oint \limits_{\partial H} \frac{-hy \mathrm{d} x}{[(x - r)^2 + h^2] \cdot [(x - r)^2 + y^2 + h^2]^{1/2}}
    \end{aligned}
\end{equation}

Finally, the geometric sensitivity \(S_{\mathrm{g}}\) can be expressed as:
\begin{equation}
S_{\mathrm{g}} = 1 - \Omega / 2\pi.
\end{equation}

\section{Appendix: Case Analysis}
\label{app:case_analysis}
Consider the union of Cap 1 and Cap \(2'\). This can be split into two cases:
\begin{enumerate}
    \item When Cap \(2'\) is fully enclosed in Cap 1, i.e. \(m + R_2 \leq R\), the union is simply Cap 1.
    \item When Cap \(2'\) extends beyond the boundry of Cap 1, i.e. \(m + R_2 > R\), the union consists of the entirety of Cap 1 plus the protruded area of Cap \(2'\).
\end{enumerate}

\begin{figure}[h]
     \centering
     \includegraphics[width=0.35\textwidth]{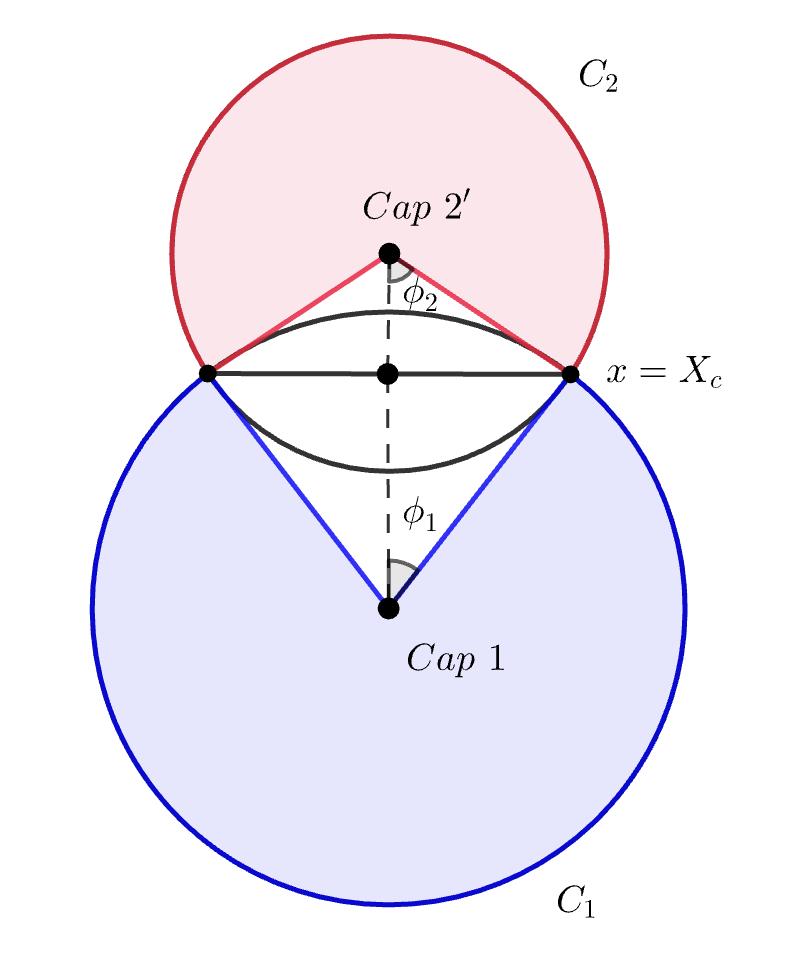}
     \caption{Diagram displaying the case when Cap 1 and Cap \(2'\) overlap. The blue arc of Cap 1 is denoted as \(C_1\) and the red arc of Cap \(2'\) is denoted as \(C_2\).}
     \label{fig:Overlap}
\end{figure}

\subsection{Case 1}
In this case, Cap \(2'\) \(\subseteq\) Cap 1, \(\partial H = x^2 + y^2 = R^2\). We can facilitate the integral by transforming \(\partial H\) into parametric form.

Take $ \theta \in [0, 2\pi), x = \cos(\theta), y = \sin(\theta) $, then

\begin{equation}
    \begin{aligned}
        \Omega &= \int_{0}^{2\pi} \frac{h}{[(\cos \theta - r)^2 + \sin^2 \theta + h^2]^{1/2}} \\
     & \cdot \frac{\sin^2 \theta \mathrm{d} \theta}{(\cos \theta - r)^2 + h^2}
    \end{aligned}
\end{equation}

\begin{equation}
    \begin{aligned}
        S_{\mathrm{g}} &= 1 - \frac{1}{2\pi} \int_{0}^{2\pi} \frac{h}{[(\cos \theta - r)^2 + \sin^2 \theta + h^2]^{1/2}} \\
     & \cdot \frac{\sin^2 \theta \mathrm{d} \theta}{(\cos \theta - r)^2 + h^2}
    \end{aligned}
\end{equation}

\subsection{Case 2}
In this case, Cap \(2'\) overlaps with Cap 1, thus we integrate over the contour of the protruded area in addition to the contour of Cap \(2'\) that is not within Cap 1. To do this, we must determine the two points of intersection between Cap 1 and Cap \(2'\). 

The X-coordinate, denoted by \(X_c\), of the common chord of two intersecting circles is given by:
\begin{equation}
    X_c = (1 - R_2^2 + m^2) / 2m
\end{equation}

With \(X_c\), we can determine the central angles, \(\phi_1\) and \(\phi_2\), of both caps for the intersection. For Cap 1, \(\cos(\phi_1) = X/R\), where \(\phi_1\) is half the central angle for Cap 1. For Cap 2’, \(\cos(\phi_2) = (m-X)/R2\), where \(\phi_2\) is half the central angle for Cap 2’. 

Define the arc of Cap 1 as $C_1$ and the arc of Cap \(2'\) as $C_2$, as illustrated by Figure~\ref{fig:Overlap}, then \(\partial H = C_1 + C_2\), and the integral becomes:
\begin{equation}
    \begin{aligned}
        \Omega &= \int_{C_1} \frac{-hy \mathrm{d} x}{[(x - r)^2 + h^2] \cdot [(x - r)^2 + y^2 + h^2]^{1/2}} \\
        & + \int_{C_2} \frac{-hy \mathrm{d} x}{[(x - r)^2 + h^2] \cdot [(x - r)^2 + y^2 + h^2]^{1/2}} \\
        &= \int_{\phi_1}^{2\pi - \phi_1} \frac{h}{[(\cos \theta - r)^2 + \sin^2 \theta + h^2]^{1/2}} \\
        & \qquad \cdot \frac{\sin^2 \theta \mathrm{d} \theta}{(\cos \theta - r)^2 + h^2} \\
        & + \int_{-\pi + \phi_2}^{\pi - \phi_2} \frac{h}{[(R_2 \cos \theta + m - r)^2 + R_2^2 \sin^2 \theta + h^2]^{1/2}} \\
        & \qquad \cdot \frac{R_2^2 \sin^2 \theta \mathrm{d} \theta}{(R_2 \cos \theta + m - r)^2 + h^2}
    \end{aligned}
\end{equation}

\begin{equation}
    \begin{aligned}
        S_{\mathrm{g}} &= 1 - \frac{\Omega}{2\pi} &\\
        &= 1 - \frac{1}{2\pi} \left( \int_{\phi_1}^{2\pi - \phi_1} \frac{h}{[(\cos \theta - r)^2 + \sin^2 \theta + h^2]^{1/2}} \right. \\
        & \qquad \qquad \cdot \frac{\sin^2 \theta \mathrm{d} \theta}{(\cos \theta - r)^2 + h^2} \\
        & \qquad + \int_{-\pi + \phi_2}^{\pi - \phi_2} \frac{h}{[(R_2 \cos \theta + m - r)^2 + R_2^2 \sin^2 \theta + h^2]^{1/2}} \\
        & \left. \qquad \qquad \cdot \frac{R_2^2 \sin^2 \theta \mathrm{d} \theta}{(R_2 \cos \theta + m - r)^2 + h^2} \right)
    \end{aligned}
\end{equation}

\bibliographystyle{unsrt}
\bibliography{PET_Sensitivity_Ref}

@article{visser2009inveon,
  title     = {Spatial Resolution and Sensitivity of the Inveon Small-Animal PET Scanner},
  author    = {Visser, Eric P. and Disselhorst, Jonathan A. and Brom, Maarten and Laverman, Peter and Gotthardt, Martin and Oyen, Wim J.G. and Boerman, Otto C.},
  journal   = {Journal of Nuclear Medicine},
  volume    = {50},
  number    = {1},
  pages     = {139--147},
  year      = {2009},
  publisher = {Society of Nuclear Medicine},
  doi       = {10.2967/jnumed.108.055152}
}

@article{jakoby2011performance,
  title        = {Physical and clinical performance of the mCT time-of-flight PET/CT scanner},
  author       = {Jakoby, B. W. and Bercier, Y. and Conti, M. and Casey, M. E. and Bendriem, B. and Townsend, D. W.},
  journal      = {Physics in Medicine and Biology},
  volume       = {56},
  number       = {8},
  pages        = {2375--2389},
  year         = {2011},
  publisher    = {IOP Publishing},
  doi          = {10.1088/0031-9155/56/8/003}
}

@article{chicheportiche2020comparison,
  title={Comparison of NEMA characterizations for Discovery MI and Discovery MI-DR TOF PET/CT systems at different sites and with other commercial PET/CT systems},
  author={Chicheportiche, Alexandre and Marciano, Rami and Orevi, Marina},
  journal={EJNMMI Physics},
  volume={7},
  number={1},
  pages={4},
  year={2020},
  publisher={Springer},
  doi={10.1186/s40658-020-0271-x},
  url={https://doi.org/10.1186/s40658-020-0271-x}
}

@article{chen2020performance,
  title={Performance characteristics of the digital uMI550 PET/CT system according to the NEMA NU2-2018 standard},
  author={Chen, Shuguang and Hu, Pengcheng and Gu, Yushen and Yu, Haojun and Shi, Hongcheng},
  journal={EJNMMI Physics},
  volume={7},
  number={1},
  pages={43},
  year={2020},
  publisher={Springer},
  doi={10.1186/s40658-020-00315-w},
  url={https://doi.org/10.1186/s40658-020-00315-w}
}

@article{bonifacio2023analytical,
  title={Analytical study of the sensitivity of cylindrical PET systems based on bulk materials and metascintillators},
  author={Bonifacio, Daniel A B and Latella, Riccardo and Murata, Helio M and Benlloch, Jose M and Gonzalez, Antonio J and Lecoq, Paul and Konstantinou, Georgios},
  journal={IEEE Transactions on Radiation and Plasma Medical Sciences},
  year={2023},
  note={Early Access},
  doi={10.1109/TRPMS.2023.3309490}
}

@manual{nema2018,
  title        = {NEMA Standards Publication NU 2-2018: Performance Measurements of Positron Emission Tomographs},
  author       = {{National Electrical Manufacturers Association (NEMA)}},
  year         = {2018},
  pages        = {17--20},
  organization = {National Electrical Manufacturers Association},
  address      = {Rosslyn, VA, USA},
  note         = {Accessed pages 17--20 for performance measurements related to PET systems.}
}

@article{amoyal2021development,
  title={Development of a hybrid gamma camera based on Timepix3 for nuclear industry applications},
  author={Amoyal, Guillaume and Schoepff, Vincent and Carrel, Fr{\'e}d{\'e}rick and Michel, Maugan and de Lanaute, N Blanc and Ang{\'e}lique, JC},
  journal={Nuclear Instruments and Methods in Physics Research Section A: Accelerators, Spectrometers, Detectors and Associated Equipment},
  volume={987},
  pages={164838},
  year={2021},
  publisher={Elsevier}
}

@incollection{Lewellen2004PETSystems,
  author    = {Thomas Lewellen and Joel Karp},
  title     = {{PET} Systems},
  booktitle = {Emission Tomography: The Fundamentals of {PET} and {SPECT}},
  editor    = {Miles N. Wernick and John N. Aarsvold},
  publisher = {Elsevier Academic Press},
  year      = {2004},
  chapter   = {10},
  pages     = {179--194},
  address   = {San Diego, CA},
  isbn      = {978-0-12-744482-6}
}

@article{Zhang2025DigitMI930,
  title     = {Performance Evaluation of New {PET/CT} {DigitMI} 930},
  author    = {Bo Zhang and Bingxuan Li and Lei Fang and Xiaoyun Zhou and Ang Li and Xuan Zhang and Yang Liu and Zhuo Wang and Chien-Min Kao and Yuqing Liu and Xiaohua Zhu and Lin Wan and Peng Xiao and Xun Chen and Hidehiro Iida and Juhani Knuuti and Qingguo Xie},
  journal   = {IEEE Transactions on Radiation and Plasma Medical Sciences},
  volume    = {9},
  number    = {5},
  pages     = {578--585},
  year      = {2025},
  month     = {May},
  publisher = {IEEE},
  doi       = {10.1109/TRPMS.2025.3526659},
  url       = {https://doi.org/10.1109/TRPMS.2025.3526659}
}

@article{SJan_2004,
doi = {10.1088/0031-9155/49/19/007},
url = {https://dx.doi.org/10.1088/0031-9155/49/19/007},
year = {2004},
month = {sep},
publisher = {},
volume = {49},
number = {19},
pages = {4543},
author = {S Jan and  G Santin and  D Strul and  S Staelens and  K Assié and  D Autret and  S Avner and  R Barbier and  M Bardiès and  P M Bloomfield and  D Brasse and  V Breton and  P Bruyndonckx and  I Buvat and  A F Chatziioannou and  Y Choi and  Y H Chung and  C Comtat and  D Donnarieix and  L Ferrer and  S J Glick and  C J Groiselle and  D Guez and  P-F Honore and  S Kerhoas-Cavata and  A S Kirov and  V Kohli and  M Koole and  M Krieguer and  D J van der Laan and  F Lamare and  G Largeron and  C Lartizien and  D Lazaro and  M C Maas and  L Maigne and  F Mayet and  F Melot and  C Merheb and  E Pennacchio and  J Perez and  U Pietrzyk and  F R Rannou and  M Rey and  D R Schaart and  C R Schmidtlein and  L Simon and  T Y Song and  J-M Vieira and  D Visvikis and  R Van de Walle and  E Wieërs and  C Morel},
title = {GATE: a simulation toolkit for PET and SPECT},
journal = {Physics in Medicine \& Biology}
}

@article{BUDINGER1998247,
title = {PET instrumentation: What are the limits?},
journal = {Seminars in Nuclear Medicine},
volume = {28},
number = {3},
pages = {247-267},
year = {1998},
note = {The Coming Age of Pet (Part 1)},
issn = {0001-2998},
doi = {https://doi.org/10.1016/S0001-2998(98)80030-5},
url = {https://www.sciencedirect.com/science/article/pii/S0001299898800305},
author = {Thomas F. Budinger}
}

@article{01630569708816786,
author = {Maryse Bourlard and Serge Nicaise},
title = {Abstract green formula and applications to boundary integral equations},
journal = {Numerical Functional Analysis and Optimization},
volume = {18},
number = {7-8},
pages = {667-689},
year = {1997},
publisher = {Taylor & Francis},
doi = {10.1080/01630569708816786},
URL = {https://doi.org/10.1080/01630569708816786},
eprint = {https://doi.org/10.1080/01630569708816786}
}

@article{ERIKSSON2007836,
title = {An investigation of sensitivity limits in PET scanners},
journal = {Nuclear Instruments and Methods in Physics Research Section A: Accelerators, Spectrometers, Detectors and Associated Equipment},
volume = {580},
number = {2},
pages = {836-842},
year = {2007},
note = {Imaging 2006},
issn = {0168-9002},
doi = {https://doi.org/10.1016/j.nima.2007.06.112},
url = {https://www.sciencedirect.com/science/article/pii/S0168900207012739},
author = {L. Eriksson and D. Townsend and M. Conti and M. Eriksson and H. Rothfuss and M. Schmand and M.E. Casey and B. Bendriem}
}

@article{nguyen2015digitalPET,
  author    = {Nghi C. Nguyen and Jose L. Vercher-Conejero and Abdus Sattar and Michael A. Miller and Piotr J. Maniawski and David W. Jordan and Raymond F. Muzic Jr. and Kuan-Hao Su and James K. O’Donnell and Peter F. Faulhaber},
  title     = {Image Quality and Diagnostic Performance of a Digital PET Prototype in Patients with Oncologic Diseases: Initial Experience and Comparison with Analog PET},
  journal   = {Journal of Nuclear Medicine},
  volume    = {56},
  number    = {9},
  pages     = {1378--1385},
  year      = {2015},
  doi       = {10.2967/jnumed.114.148338},
  publisher = {Society of Nuclear Medicine and Molecular Imaging}
}

@article{Calderon2023,
  author    = {Eduardo Calderón and Fabian P. Schmidt and Wenhong Lan and Salvador Castaneda-Vega and Andreas S. Brendlin and Nils F. Trautwein and Helmut Dittmann and Christian la Fougère and Lena Sophie Kiefer},
  title     = {Image Quality and Quantitative PET Parameters of Low-Dose [18F]FDG PET in a Long Axial Field-of-View PET/CT Scanner},
  journal   = {Diagnostics},
  year      = {2023},
  volume    = {13},
  number    = {20},
  pages     = {3240},
  doi       = {10.3390/diagnostics13203240},
  url       = {https://www.mdpi.com/2075-4418/13/20/3240},
  publisher = {MDPI},
}

@article{Madsen2004,
  author    = {Mark T. Madsen},
  title     = {Advances in PET Imaging},
  journal   = {Correspondence Continuing Education Courses for Nuclear Pharmacists and Nuclear Medicine Professionals},
  volume    = {10},
  number    = {5},
  year      = {2004},
  publisher = {University of New Mexico Health Sciences Center, College of Pharmacy},
  note      = {Accredited by the American Council on Pharmaceutical Education, Program No. 039-000-02-005-H04, 3.5 Contact Hours, 0.35 CEUs.}
}

@article{Li2017PET,
  author    = {Zheng Li and Anisha A. Gupte and Anjun Zhang and Dale J. Hamilton},
  title     = {PET Imaging and Its Application in Cardiovascular Diseases},
  journal   = {Methodist DeBakey Cardiovascular Journal},
  volume    = {13},
  number    = {1},
  pages     = {29-32},
  year      = {2017},
  publisher = {Houston Methodist Research Institute},
}

@article{Farwell2014PET,
  author    = {Michael D. Farwell and Daniel A. Pryma and David A. Mankoff},
  title     = {PET/CT Imaging in Cancer: Current Applications and Future Directions},
  journal   = {Cancer},
  volume    = {120},
  number    = {22},
  pages     = {3433-3445},
  year      = {2014},
  publisher = {American Cancer Society},
  doi       = {10.1002/cncr.28860},
  url       = {https://doi.org/10.1002/cncr.28860}
}

@article{Politis2012PET,
  author    = {Marios Politis and Paola Piccini},
  title     = {Positron emission tomography imaging in neurological disorders},
  journal   = {Journal of Neurology},
  volume    = {259},
  pages     = {1769-1780},
  year      = {2012},
  publisher = {Springer-Verlag},
  doi       = {10.1007/s00415-012-6428-3},
  url       = {https://doi.org/10.1007/s00415-012-6428-3}
}

@article{PerezBenito2023,
  author    = {D. Perez-Benito and R. Chil and L. A. Hidalgo-Torres and J. J. Vaquero},
  title     = {Scintillator Geometrical Considerations for Detectors Based on Hexagonal SiPMs},
  journal   = {IEEE Transactions on Radiation and Plasma Medical Sciences},
  volume    = {7},
  number    = {7},
  pages     = {684-691},
  year      = {2023},
  month     = {September},
  doi       = {10.1109/TRPMS.2023.3279654},
  publisher = {IEEE},
  url       = {https://doi.org/10.1109/TRPMS.2023.3279654}
}

@article{Ficke1996SpheroidPET,
  author = {David C. Ficke and John T. Hood and Michel M. Ter-Pogossian},
  title = {A Spheroid Positron Emission Tomograph for Brain Imaging: A Feasibility Study},
  journal = {Journal of Nuclear Medicine},
  volume = {37},
  number = {7},
  pages = {1219-1225},
  year = {1996},
  publisher = {Society of Nuclear Medicine},
  url = {https://jnm.snmjournals.org/content/37/7/1219},
}

@incollection{Vallabhajosula2023,
  author    = {S. Vallabhajosula},
  title     = {Radioactivity Detection: PET and SPECT Scanners},
  booktitle = {Molecular Imaging and Targeted Therapy},
  publisher = {Springer},
  year      = {2023},
  chapter   = {5},
  doi       = {10.1007/978-3-031-23205-3_5},
  url       = {https://link.springer.com/book/10.1007/978-3-031-23205-3}
}

@book{CherryDahlbomPETPhysics,
  author = {Simon R. Cherry and Magnus Dahlbom},
  title = {PET: Physics, Instrumentation, and Scanners},
  publisher = {Unknown Publisher},
  year = {Unknown Year},
  note = {Comprehensive guide on the physics and instrumentation of positron emission tomography (PET)},
}

@phdthesis{Camborde2001,
  author    = {M. L. A. Camborde},
  title     = {Detection to Improve the Quality of PET Imaging},
  school    = {McGill University},
  year      = {2001},
  url       = {https://www.collectionscanada.gc.ca/obj/s4/f2/dsk4/etd/MQ78840.PDF},
  note      = {Accessed: 2024-02-21}
}

@ARTICLE{9353691,
  author={Gonzalez-Montoro, Andrea and Gonzalez, Antonio J. and Pourashraf, Shirin and Miyaoka, Robert S. and Bruyndonckx, Peter and Chinn, Garry and Pierce, Larry A. and Levin, Craig S.},
  journal={IEEE Transactions on Radiation and Plasma Medical Sciences}, 
  title={Evolution of PET Detectors and Event Positioning Algorithms Using Monolithic Scintillation Crystals}, 
  year={2021},
  volume={5},
  number={3},
  pages={282-305},
  doi={10.1109/TRPMS.2021.3059181}}
\end{document}